\documentclass[12pt]{article}
\usepackage{epsf}
\usepackage{amsmath, amssymb}
\usepackage{graphicx}
\usepackage{here}
\usepackage{subfigure}

\def\be{\begin{equation}}
\def\ee{\end{equation}}
\def\bea{\begin{eqnarray}}
\def\eea{\end{eqnarray}}
\def\fr{\frac}
\def\der{\partial}

\def\({\left(}
\def\){\right)}
\def\<{\langle}
\def\>{\rangle}
\newcommand{\nn}{\nonumber \\}

\begin{document}

\renewcommand{\theequation}{\thesection.\arabic{equation}}
\renewcommand{\thefootnote}{\fnsymbol{footnote}}
\setcounter{footnote}{0}
\begin{titlepage}

\def\thefootnote{\fnsymbol{footnote}}

\begin{center}

\hfill UT-11-39\\
\hfill IPMU-11-0190\\
\hfill November, 2011\\

\vskip .75in

{\Large \bf 

Boltzmann equation for non-equilibrium particles and its
application to non-thermal dark matter production

}

\vskip .75in

{\large
Koichi Hamaguchi$^{(a,b)}$, Takeo Moroi$^{(a,b)}$ and Kyohei Mukaida$^{(a)}$
}

\vskip 0.25in

{\em
$^a$ Department of Physics, University of Tokyo,
Tokyo 113-0033, Japan\\
$^b$ Institute for the Physics and Mathematics of the Universe, 
University of Tokyo, Kashiwa 277-8568, Japan
}

\end{center}
\vskip .5in

\begin{abstract}

  We consider a scalar field (called $\phi$) which is very weakly
  coupled to thermal bath, and study the evolution of its number
  density.  We use the Boltzmann equation derived from the
  Kadanoff-Baym equations, assuming that the degrees of freedom in the
  thermal bath are well described as ``quasi-particles.''  When the
  widths of quasi-particles are negligible, the evolution of the
  number density of $\phi$ is well governed by a simple Boltzmann
  equation, which contains production rates and distribution functions
  both evaluated with dispersion relations of quasi-particles with
  thermal masses.  We pay particular attention to the case that dark
  matter is non-thermally produced by the decay of particles in
  thermal bath, to which the above mentioned formalism is applicable.
  When the effects of thermal bath are properly included, the relic
  abundance of dark matter may change by $O(10-100\ \%)$ compared to
  the result without taking account of thermal effects.

\end{abstract}

\end{titlepage}

\renewcommand{\thepage}{\arabic{page}}
\setcounter{page}{1}
\renewcommand{\thefootnote}{\#\arabic{footnote}}
\setcounter{footnote}{0}

\section{Introduction}
\setcounter{equation}{0}

In particle cosmology, it is inevitable to consider the behavior of
quantum fields (or, in other words, particles) in thermal bath because
the universe was filled with hot plasma in the early epoch.  The
detailed thermal effects depend on how the particle of our interest,
which we call $\phi$, interacts with degrees of freedom in thermal
bath.  Importantly, even if $\phi$ is so weakly interacting that it is
not in thermal equilibrium,
there can be non-negligible thermal effects on its dynamics.
This is because the interaction rate of $\phi$ surrounded by
thermal bath depends on the properties (in particular, the dispersion
relation) of the degrees of freedom in thermal bath, which can of
course be significantly affected by thermal effects.

One important example of such a very weakly interacting particle is
non-thermal dark matter which is produced by the decay of particles in
thermal bath.  Although the existence of dark matter is strongly
suggested by various cosmological observations~\cite{Nakamura:2010zzi},
particle-physics properties of dark matter, as well as its production
mechanism in the early universe, have not been understood yet.
Various particle physics models including dark matter candidate have
been proposed so far, like supersymmetric models, 
universal extra dimension models, and so on~\cite{Bertone:2004pz}.  In
the future, it is hoped that those models are tested by high energy
experiments as well as by cosmological observations.  In particular,
the candidate of the dark matter particle may be discovered and
studied by the LHC experiment as well as future linear colliders,
based on which a large class of dark matter models are discriminated.
For this program, precise theoretical calculation of the relic
abundance of the dark matter candidate should be performed by using
information about newly discovered particles \cite{DM_collider}.
The present dark matter density is
very accurately determined by the WMAP collaboration as
\cite{Komatsu:2010fb}:
\begin{eqnarray}
  \Omega_c h^2=0.1126 \pm 0.0036,
\end{eqnarray}
with $h$ being the Hubble constant in units of $100$ km/sec/Mpc, so the dark
matter abundance is now known with $O(1\ \%)$ accuracy.  Thus, it is
desirable to establish methods of calculating the dark matter density
at the same level of accuracy.  For this purpose, detailed
understanding of thermal effects on the dark matter production process
is required.

In the present study, we pay particular attention to the case that
dark matter particle is non-thermally produced by the decay of heavier
particles in thermal bath.
There are many examples of such non-thermally produced dark matter,
like gravitino~\cite{Moroi:1993mb}, axino~\cite{Covi:2001nw}, a
singlet field \cite{McDonald:2001vt}, the right handed
sneutrino~\cite{Asaka:2005cn+2006}, and more generally, the recently
proposed ``freeze-in'' particles~\cite{Hall:2009bx}.  In such a
scenario, the relic density of dark matter is determined at the cosmic
temperature comparable to the mass of decaying particle and is
insensitive to the thermal history before that.
In the previous studies, the production rate of
dark matter has been calculated by using 
the decay rates of particles
estimated in vacuum.  However, in the actual situation, 
the particles decay in the thermal bath, so the production rate
taking account of the thermal effects should be properly used in the
calculation of the dark matter density.  
As we will see, the thermal effects may significantly change the resultant 
abundance of dark matter.

In this paper, we raise the question how important the thermal effects
are in the production process of non-thermal dark matter.
To answer this question, we first study the properties of Boltzmann equation derived from the
Kadanoff-Baym equations~\cite{Kadanoff:1962}
under the assumption 
that the production of dark matter does not affect the thermal bath. 
In particular, 
unless the dark matter production is 
almost kinematically blocked by thermal masses
during the time when the production of dark matter is most effective, 
the full Boltzmann equation can be reduced to a simplified form which has the
same structure as the conventional Boltzmann equation\footnote{
In this paper, the ``conventional'' Boltzmann equation refers to
the Boltzmann equation evaluated with zero-temperature dispersion 
relations~\cite{KolbTurner}.
} but constructed
with ``thermal masses'' of particles in thermal bath.  (As we will
see, such a simplified Boltzmann equation is obtained by taking the
``zero-width approximation'' of particles in thermal bath.)  We
evaluate the relic density of dark matter by solving (i) the full, 
(ii) zero-width
approximated, and (iii) conventional Boltzmann
equations.  Comparing the three results, we discuss how important the
thermal effects are and when the zero-width approximation breaks down.
We will see that the dark matter abundance may be reduced by $O(10-100\
\%)$ compared with the result of calculation where the thermal effects
are neglected.

The organization of this paper is as follows.  In Section
\ref{sec:formalism}, the relevant formulae to study the evolution of
the number density of non-equilibrium particles are summarized.  In
particular, properties of the Boltzmann equation to be solved are
discussed.  Then, in Section \ref{sec:dm}, we apply the formalism to
the non-thermal dark matter production process.  We numerically solve
the Boltzmann equation and discuss how important the thermal effects
are.  Section \ref{sec:conclusions} is devoted to conclusions and
discussion.

\section{Formalism}
\label{sec:formalism}
\setcounter{equation}{0}

First, let us introduce the formulae and equations relevant for our
analysis.  Although many of them can be found in literature (see, for
instance, \cite{Kadanoff:1962, Anisimov:2008dz, Kubo:1957mj+x,
  Dolan:1973qd, Hosoya:1983ke+x, Yokoyama:2005dv, Drewes:2010pf,
  Literatures1, Literatures2, Literatures3, Literatures4,
  Literatures5}), we summarize the relevant equations to make this
paper self-contained for the sake of readers.  We assume that the
thermal bath have a common temperature $T$, which have a large degrees
of freedom, and the back reaction to the thermal bath from the
production of $\phi$ is negligible.

In this paper, we study the evolution of the number density of a
scalar field $\phi$ coupled to scalar fields in thermal bath, which
are denoted as $\chi_i$.  We introduce the
interaction of the following form:
\begin{eqnarray}
  {\cal L}_{\rm int} = g \phi \prod_{i=1}^n \chi_i
  = g \phi {\cal O}[\chi_0,\chi_1, \cdots],
  \label{L_int}
\end{eqnarray}
where $g$ is a coupling constant and, for the convenience of the
following discussion, we introduced the operator ${\cal
  O}\equiv\prod_i \chi_i$.  The interaction of $\phi$ is assumed to be
extremely small, i.e., $gm_\phi^{n-3} \ll 1$, where $m_\phi$ is the
mass of $\phi$.\footnote{
   In the case of freeze-in dark matter
   with a weak scale mass, typically a coupling of $O(10^{-13})$ 
   is necessary to obtain the correct dark matter abundance~\cite{Asaka:2005cn+2006,Hall:2009bx}.
}
In the
following, we study the effects which are leading order in $g$.  For
simplicity, we consider the case that ${\cal O}$ is given by a product
of scalar fields.  However, the extension of the formalism to the case
that ${\cal O}$ includes derivatives of scalar fields is
straightforward.

The evolution of the number density of $\phi$ in the early universe is
governed by two effects: one is the production of $\phi$ due to the
decay and scattering processes and the other is the cosmic expansion.  We discuss these
effects separately.

Because of the weakness
of the interaction, $\phi$ can be regarded as (almost) free
particle,\footnote
{We assume that the self-interaction of $\phi$ is absent or weak enough to
  be neglected. 
  We also assume that the thermal effects on the expectation value of $\phi$ is negligible,
  which is the case in the example in Section \ref{sec:dm}.
  }
and the number density operator is given by
\begin{eqnarray}
  \hat{N}_{{\bf k}} (t) = 
  \frac{1}{2\omega_{\bf k}}
  : \left[
    \dot{\hat{\phi}}(t;{\bf k}) \dot{\hat{\phi}}(t;{-\bf k}) 
    + \omega_{\bf k}^2 \hat{\phi}(t;{\bf k}) \hat{\phi}(t;{-\bf k})
  \right] :,
  \label{hatN_k}
\end{eqnarray}
where the ``dot'' denotes the derivative with respect to time,
$:\cdots:$ is the normal ordering and 
\begin{eqnarray}
  \omega_{\bf k}\equiv\sqrt{{\bf k}^2 + m_\phi^2}.
\end{eqnarray}
In addition,
\begin{eqnarray}
  \hat{\phi} (t;{\bf k}) \equiv L^{-3/2}
  \int d^3 x\, e^{-i{\bf k}\cdot{\bf x}} \hat{\phi} (t, {\bf x}),
\end{eqnarray}
where $\hat{\phi}$ is the field operator for $\phi$ and $L^3$ is the
volume of the system (where we have adopted box normalization).  For
the later convenience, the number density for each momentum eigenstate
is defined.
Then, the expectation value of the number density of $\phi$ is
obtained by using the density matrix $\hat{\rho}$:
\begin{eqnarray}
  N_{{\bf k}} (t) \equiv 
  \langle \hat{N}_{{\bf k}} (t) \rangle,
  \label{<N_k>}
\end{eqnarray}
where, for an operator $\hat{A}$,
\begin{eqnarray}
  \langle \hat{A} \rangle \equiv
  {\rm tr}[\hat{\rho} \hat{A}].
\end{eqnarray}

The $\chi$ sector is in the thermal bath while $\phi$ is always out of
thermal equilibrium.  In particular, we are interested in the case
that $\phi$ is initially absent in the system.  Thus, we assume that
the initial density matrix (at $t=t_i$) is given by the direct product
of density matrices of two sectors:
\begin{eqnarray}
  \hat{\rho} = \hat{\rho}_{\phi, i} \otimes \hat{\rho}_{\chi}.
\end{eqnarray}
We set $t_i=0$ without loss of generality.  The $\chi$ sector is in
the thermal bath (with the temperature $T$), so $\hat{\rho}_{\chi}$ is
given by
\begin{eqnarray}
  \hat{\rho}_{\chi} = e^{-\hat{H}_\chi/T},
\end{eqnarray}
with $\hat{H}_\chi$ being the Hamiltonian for the $\chi$ sector.  On
the contrary, $\hat{\rho}_{\phi,i}$ determines the initial
distribution of $\phi$.  We assume that it has translational invariance,
i.e., $[\hat{\bf P}, \hat{\rho}_{\phi,i}]=0$, where $\hat{\bf P}$ is the
momentum operator.  Furthermore, we assume
$\langle\hat{\phi}(t=0)\rangle=\langle\dot{\hat{\phi}}(t=0)\rangle=0$.

In order to calculate the evolution of $N_{\bf k}(t)$, we define the
Hadamard propagator and the Jordan propagator:
\begin{eqnarray}
  G_{\rm H}^\phi (t,t';{\bf k}) &\equiv&
  \langle \hat{\phi}(t;{\bf k}) \hat{\phi}(t';-{\bf k}) \rangle
  + \langle \hat{\phi}(t';-{\bf k}) \hat{\phi}(t;{\bf k}) \rangle, \\ 
  G_{\rm J}^\phi (t,t';{\bf k}) &\equiv&
  \langle \hat{\phi}(t;{\bf k}) \hat{\phi}(t';-{\bf k}) \rangle
  - \langle \hat{\phi}(t';-{\bf k}) \hat{\phi}(t;{\bf k}) \rangle.
\end{eqnarray}
As can be seen from Eqs.\ \eqref{hatN_k} and \eqref{<N_k>}, the
expectation value of the number density is given by
\begin{eqnarray}
  N_{{\bf k}} (t) \equiv 
  \frac{1}{4\omega_{\bf k}}
  \left[
    (\partial_t \partial_{t'} + \omega_{\bf k}^2)
    G_{\rm H}^\phi (t,t';{\bf k})
  \right]_{t'\rightarrow t} - C_{{\bf k}},
\end{eqnarray}
where $C_{{\bf k}}$ is normal-ordering constant.  In the weak coupling
limit, $C_{{\bf k}}=\frac{1}{2}$.  The Hadamard propagator and the
Jordan propagator satisfy the following equations, namely the 
Kadanoff-Baym equations~\cite{Kadanoff:1962, Anisimov:2008dz}
\begin{eqnarray}
  \( \partial_t^2 + \omega_{\bf k}^2 \) 
  G_{\rm J}^\phi(t,t';{\bf k}) 
  &=&- \int_{t'}^{t} d\tau \Pi_{\rm ret}^\phi(t - \tau ; {\bf k})
  G_{\rm J}^\phi (\tau,t';{\bf k}),
  \label{eq:KB2}
  \\
  \( \partial_t^2 + \omega_{\bf k}^2 \) 
  G_{\rm H}^\phi(t,t';{\bf k}) 
  &=&- \int_0^{t} d\tau \Pi_{\rm ret}^\phi(t - \tau ; {\bf k})
  G_{\rm H}^\phi (\tau,t';{\bf k}) 
  \nonumber \\
  &&-i \int^{t'}_0 d\tau 
  \Pi_H^\phi(t-\tau ;{\bf k})G_{\rm J}^\phi(\tau,t';{\bf k}), 
  \label{eq:KB1}
\end{eqnarray}
where
\begin{eqnarray}
  \Pi^\phi_{\rm ret}(t;{\bf k}) &=&
  -i\theta(t) 
  \left(\Pi^\phi_>(t;{\bf k}) - \Pi^\phi_<(t;{\bf k})\right),
  \label{eq:Pi_ret}
  \\
  \Pi^\phi_H(t;{\bf k}) 
  &=& \Pi^\phi_>(t;{\bf k}) + \Pi^\phi_<(t;{\bf k}).
\end{eqnarray}
Here, at the leading order (i.e, $O(g^2)$),
\begin{eqnarray}
  \Pi^\phi_>(t;{\bf k}) &=& 
  \frac{g^2}{{\rm tr} [e^{-\hat{H}_\chi/T}]}
  {\rm tr}\left[ e^{-\hat{H}_\chi/T} 
    \hat{\cal O}(t;{\bf k})\hat{\cal O} (0;-{\bf k})\right]\,,
  \label{eq:Pi>}
  \\
  \Pi^\phi_<(t;{\bf k}) &=& 
  \frac{g^2}{{\rm tr} [e^{-\hat{H}_\chi/T}]}
  {\rm tr}\left[e^{-\hat{H}_\chi/T} 
    \hat{\cal O} (0;-{\bf k}) \hat{\cal O}(t;{\bf k})\right]\,,
  \label{eq:Pi<}
\end{eqnarray}
with
\begin{eqnarray}
  \hat{\cal O}(t;{\bf k}) &=& L^{-3/2}
  \int d^3 x e^{-i{\bf k}\cdot{\bf x}} \hat{\cal O} (t,{\bf x}).
\end{eqnarray}
Let us define the Fourier transformations
\begin{eqnarray}
  \Pi^\phi_X(\omega,{\bf k}) = \int dt e^{i\omega t} \Pi^\phi_X(t;{\bf k}),
\end{eqnarray}
with $\Pi_X=\Pi_{\rm ret}$, $\Pi_H$, $\Pi_<$, and $\Pi_>$.
Then, from Eq.~(\ref{eq:Pi_ret}), 
\begin{eqnarray}
  \Pi^\phi_{\rm ret}(\omega, {\bf k}) &=&
  \int \fr{d \omega'}{2\pi} 
  \frac{\Pi_{>}^\phi (\omega',{\bf k})-\Pi_{<}^\phi (\omega',{\bf k})}
  {\omega-\omega'+i0}.
\end{eqnarray}
By using the
relation $(\omega+i0)^{-1}={\cal P}\omega^{-1}-i\pi\delta(\omega)$
(with ${\cal P}$ denoting the principal value), we obtain
\begin{eqnarray}
  \Im\Pi^\phi_{\rm ret} (\omega,{\bf k}) = 
  -\fr{1}{2} 
  \left[ \Pi_{>}^\phi (\omega,{\bf k})-\Pi_{<}^\phi (\omega,{\bf k}) \right],
  \label{eq:ImPiret_Pi><}
\end{eqnarray}
and
\begin{eqnarray}
  \Pi^\phi_H(\omega, {\bf k}) =
  -2
  \coth \left( \frac{\omega}{2T} \right)
  \Im \Pi^\phi_{\rm ret}(\omega,{\bf k}),
  \label{eq:PiH_Pi><}
\end{eqnarray}
where we have used the so-called Kubo-Martin-Schwinger (KMS) relation~\cite{Kubo:1957mj+x} $\Pi_>(\omega,{\bf k})=\exp(\omega/T)\Pi_<(\omega,{\bf k})$
in deriving Eq.\ \eqref{eq:PiH_Pi><}.

Using the fact that the Jordan propagator is time translational
invariant within our setup,\footnote
{This is because, under no self interactions of $\phi$ and truncating
  the perturbative expansion at ${\cal O}(g^2)$, the spectrum of
  $\phi$ is determined only by the thermal bath regardless of the
  number density of $\phi$.  The time translational invariance of the
  Jordan propagator is not a general property. See also \cite{Anisimov:2008dz}.}
 $G_{\rm J}^\phi (t,t';{\bf k}) = G_{\rm J}^\phi
(t-t',0;{\bf k})$, it can be expressed in terms of the spectral density $\rho_\phi$
as\footnote
{The spectral density $\rho_{\phi}$ should not be confused with the
  density matrix $\hat{\rho}_{\phi,i}$.}
\begin{eqnarray}
  G_{\rm J}^\phi(t;{\bf k})\equiv G_{\rm J}^\phi(t,0;{\bf k}) \equiv
  \int \fr{d\omega}{2\pi} e^{-i\omega t}\rho_\phi (\omega,{\bf k}),
  \label{eq:GJ}
\end{eqnarray} 
and the solution to Eq.\ \eqref{eq:KB2} is given by
\begin{eqnarray}
  \rho_\phi (\omega,{\bf k}) = 
  \frac{-2\Im \Pi^\phi_{\rm ret}(\omega,{\bf k})}
  {[\omega^2 - \omega_{\bf k}^2 
    - \Re \Pi^\phi_{\rm ret}(\omega,{\bf k})]^2
    + [\Im \Pi^\phi_{\rm ret}(\omega,{\bf k})]^2}.
\end{eqnarray}
We note that the initial condition is given by $G_{\rm J}^\phi (0;{\bf
  k})=0$ and $\der_t G_{\rm J}^\phi(0;{\bf k})=-i$ because of  
equal time commutation relations.

With the initial condition of the Jordan propagator, the Hadamard
propagator satisfying the Kadanoff-Baym equations \eqref{eq:KB2} and
\eqref{eq:KB1} is obtained as follows \cite{Anisimov:2008dz}:
\begin{eqnarray}
  G_{\rm H}^\phi(t,t';{\bf k}) 
  = G_{\rm H}^{\rm hom}(t,t';{\bf k}) 
  + \int^{t}_0 dt_1 \int^{t'}_0  dt_2 
  G_{\rm J}^\phi(t-t_1;{\bf k})\Pi_H^\phi(t_1-t_2;k)G_{\rm J}^\phi(t_2-t';{\bf k}),
\end{eqnarray}
where the homogeneous solution is given by
\begin{eqnarray}
  G_{\rm H}^{\rm hom}(t,t';{\bf k})
  &=&
  -\left. G_{\rm H}^\phi(s,s';{\bf k}) \right|_{s,s'=0}   
  \der_t \der_{t'}G_{\rm J}^\phi(t;{\bf k}) G_{\rm J}^\phi(t';{\bf k}) 
  \nn
  &&   -
  \left. \der_s G_{\rm H}^\phi(s,s';{\bf k}) \right|_{s,s'=0} 
  \left(\der_t + \der_{t'}\right)
  G_{\rm J}^\phi(t;{\bf k}) G_{\rm J}^\phi(t';{\bf k}) 
  \nn
  &&-
  \left. \der_s \der_{s'} G_{\rm H}^\phi(s,s';{\bf k})\right|_{s,s'=0}   
  G_{\rm J}^\phi(t;{\bf k}) G_{\rm J}^\phi(t';{\bf k})\,.
\end{eqnarray}

In the calculation of the production rate of $\phi$ in thermal bath,
the most important effect is the shift of the pole of the spectral
density because its imaginary part gives the production rate.  Here,
we are interested in the case that the interaction of $\phi$ is so
small that $\omega_{\bf k}^2\gg|\Pi^\phi_{\rm ret}|$.  Then, the
spectral density can be well approximated by the Breit-Wigner form:
\begin{eqnarray}
  \rho_\phi^{\rm (BW)} (\omega,{\bf k}) = 
  \frac{2\omega \Gamma_{\phi}({\bf k})}
  {(\omega^2 - \omega_{\bf k}^2)^2 + (\omega \Gamma_{\phi}({\bf k}))^2},
  \label{rho_phi(BW)}
\end{eqnarray}
where 
\begin{eqnarray}
  \Gamma_\phi ({\bf k}) \equiv 
  -\frac{\Im \Pi^\phi_{\rm ret}(\omega_{\bf k},{\bf k})}{\omega_{\bf k}}.
  \label{Gamma_k}
\end{eqnarray}
Although $\Gamma_\phi$ has the argument ${\bf k}$ in our expression,
it depends only on $|{\bf k}|$ because of the rotational invariance of
the thermal bath.  Here, we neglected the correction to the real part
of the pole, which is expected to be irrelevant.  In addition, notice
that $\Gamma_\phi ({\bf k})$ is of $O(g^2)$, and is much smaller than
$\omega_{\bf k}$.  Then, from Eq.~(\ref{eq:GJ}), the Jordan propagator
(for $t\geq 0$) is well approximated as
\begin{eqnarray}
  \left. iG_{\rm J}^\phi (t;{\bf k}) \right|_{t\geq 0} = 
  \frac{\sin \omega_{\bf k} t}{\omega_{\bf k}} 
  e^{-\Gamma_\phi ({\bf k}) t / 2},
  \label{f_k(BW)}
\end{eqnarray}
resulting in the following expression for the expectation value of the
number density defined in Eq.\ \eqref{<N_k>},
at leading order in $\Gamma_\phi/\omega_{\bf k}$,
\begin{eqnarray}
  N_{{\bf k}}^{\rm (BW)} (t) &=& f_{\rm B} (\omega_{\bf k}) 
  \left(1 - e^{-\Gamma_\phi ({\bf k}) t} \right) \nn
  &&+ \frac{1}{4\omega_{\bf k}} 
  \left[
    \left. G_{\rm H}^\phi(s,s';{\bf k}) \right|_{s,s'=0}
     \omega_{\bf k}^2
    + \left. \der_s \der_{s'} G_{\rm H}^\phi(s,s';{\bf k}) \right|_{s,s'=0}
    - 2\omega_{\bf k}
  \right] 
  e^{-\Gamma_\phi ({\bf k}) t},
  \nonumber \\
\end{eqnarray}
with
\begin{eqnarray}
  f_{\rm B} (\omega) = \frac{1}{e^{\omega/T}-1}.
\end{eqnarray}
Equivalently, irrespective of the initial condition, one finds
\begin{eqnarray}
  \dot{N}_{{\bf k}}^{\rm (Coll)} = 
  \Gamma_\phi ({\bf k}) 
  \left[ f_{\rm B} (\omega_{\bf k}) - N_{{\bf k}} \right],
  \label{Boltzmann_k}
\end{eqnarray}
from which we can see that $\Gamma_\phi ({\bf k})$ can be regarded as the
production rate of $\phi$ due to the decays and scatterings of particles in thermal
bath.  (Here, the superscript ``(Coll)'' implies that this is the
collision term in the Boltzmann equation.)

As we have mentioned, there is another effect
on the evolution of the $\phi$'s number density, which is
the expansion of the universe.  Effect of the cosmic expansion can be
easily evaluated by taking into account the red-shift of the momentum
and we obtain
\begin{eqnarray}
  \dot{N}_{{\bf k}}^{\rm (Exp)} = 
  H {\bf k}\cdot \frac{\partial N_{{\bf k}}}{\partial {\bf k}}, 
\end{eqnarray}
where the superscript ``(Exp)'' is for cosmic expansion,
and $H$ denotes Hubble parameter of the expanding universe.
We assume that the energy density of the thermal bath 
is much larger than that of $\phi$, and that the Hubble
parameter depends only on the temperature $T$.

Combining two effects, the Boltzmann equation to be solved is
\begin{eqnarray}
  \dot{N_{{\bf k}}} - 
  H {\bf k} \cdot\frac{\partial N_{{\bf k}}}{\partial {\bf k}}
  = 
  \Gamma_\phi ({\bf k};T) 
  \left[ f_{\rm B} (\omega_{\bf k}; T) - N_{{\bf k}} \right],
  \label{Boltzmann_k_exp}
\end{eqnarray}
or, for the total number density
\begin{eqnarray}
  n_\phi \equiv \int \frac{d^3 k}{(2\pi)^3} N_{{\bf k}},
\end{eqnarray}
the Boltzmann equation is given by
\begin{eqnarray}
  \frac{d n_\phi}{d t} + 3 H n_\phi &=& 
  \dot{n}_\phi^{\rm (Coll)}
  \nonumber\\
  &\equiv& 
  \int \frac{d^3 k}{(2\pi)^3}
  \Gamma_\phi ({\bf k};T) 
  \left[ f_{\rm B} (\omega_{\bf k}; T) - N_{{\bf k}} \right].
  \label{eq:Ncoll}
\end{eqnarray}
In the above equations, we explicitly show that $\Gamma_\phi$ and
$f_{\rm B}$ depend on $T$, which should be identified with the cosmic
temperature.  Thus, the most important quantity to study the
evolution of the abundance of $\phi$ is the production rate $\Gamma_\phi({\bf
  k};T)$ given in Eq.\ \eqref{Gamma_k} or, equivalently,
$\Im\Pi^\phi_{\rm ret}$ given in Eq.\ \eqref{eq:ImPiret_Pi><}.  With
the operator ${\cal O}$ given in Eq.\ \eqref{L_int}, the leading 
contribution to $\Im\Pi^\phi_{\rm ret}$ is given by
\begin{eqnarray}
  \Im\Pi^\phi_{\rm ret} (\omega_{\bf k}, {\bf k}) &=&
  \frac{g^2}{2} \int 
  \left[ \prod_i \frac{d^4 p_i}{(2\pi)^4} G^{\chi_i}_<(p_i^0,{\bf p}_i) \right]
  (2\pi)^4 
  \delta (\omega_{\bf k} - \mbox{$\sum_i$} p_i^0) 
  \delta^{(3)} ({\bf k}- \mbox{$\sum_i$} {\bf p}_i) 
  \nonumber \\ &&
  - \frac{g^2}{2} \int 
  \left[ \prod_i \frac{d^4 p_i}{(2\pi)^4} G^{\chi_i}_>(p_i^0,{\bf p}_i) \right]
  (2\pi)^4 
  \delta (\omega_{\bf k} - \mbox{$\sum_i$} p_i^0) 
  \delta^{(3)} ({\bf k}- \mbox{$\sum_i$} {\bf p}_i).
  \nonumber \\
\end{eqnarray}
Here, all the $\chi_i$-fields are taken to be independent.  If there
exist identical fields in $\{\chi_i\}$ in Eq.\ \eqref{L_int}, a 
symmetry factor is needed.  In the above expression, the functions
$G^{\chi_i}_>$ and $G^{\chi_i}_<$ are defined as
\begin{eqnarray}
  G^{\chi_i}_>(p^0,{\bf p}) &=& 
  \frac{1}{{\rm tr} [e^{-H_\chi/T}]}
  \int d^4 x e^{ip^0 x^0 - i {\bf p}\cdot {\bf x}}
  {\rm tr} [e^{-H_\chi/T}\chi_i(x^0, {\bf x})\chi_i(0,0)],
  \\
  G^{\chi_i}_<(p^0,{\bf k}) &=& 
  \frac{1}{{\rm tr} [e^{-H_\chi/T}]}
  \int d^4 x e^{ip^0 x^0 - i {\bf p}\cdot {\bf x}}
  {\rm tr} [e^{-H_\chi/T}\chi_i(0,0)\chi_i(x^0, {\bf x})].
\end{eqnarray}
These functions satisfy the KMS relation $G_>^{\chi_i}(p^0,{\bf
  p})=e^{p^0/T}G_<^{\chi_i}(p^0,{\bf p})$, whereas their difference,
the Jordan propagator, is expressed in terms of the spectral density:
\begin{eqnarray}
  G_J^{\chi_i}(p^0,{\bf p}) = 
  G^{\chi_i}_>(p^0,{\bf p}) -G^{\chi_i}_<(p^0,{\bf p}) = 
  \rho_{\chi_i}(p^0,{\bf p}).
\end{eqnarray}
Thus, they are given by
\begin{eqnarray}
  G^{\chi_i}_>(p^0,{\bf p}) &=& 
  (f_{\rm B} (p^0) + 1) \rho_{\chi_i}(p^0,{\bf p}),
  \\
  G^{\chi_i}_<(p^0,{\bf p}) &=& 
  f_{\rm B} (p^0) \rho_{\chi_i}(p^0,{\bf p}).
\end{eqnarray}

If the quasi-particle picture is applicable to $\chi_i$, which we
assume in the following analysis, the spectral density can be
approximated by the Breit-Wigner form:
\begin{eqnarray}
  \rho_{\chi_i}^{\rm (BW)} (p^0,{\bf p})=
  \frac{2p^0 \Gamma_{\chi_i}({\bf p};T)}
  {({p^0}^2 - \Omega_{\chi_i}^2({\bf p};T))^2 
    + (p^0 \Gamma_{\chi_i}({\bf p};T))^2},
\end{eqnarray}
where $\Omega_{\chi_i}$ and $\Gamma_{\chi_i}$ are real and imaginary
parts of the pole of the propagator, and are related to the real and
imaginary parts of the self energy of $\chi_i$.  Contrary to the case
of $\phi$, we need to take account of the shirt of the pole for
particles which are thermalized.  Thus, $\Omega_{\chi_i}$ may
significantly deviate from the frequency satisfying the on-shell
condition in the vacuum.  In a large class of models, including the case
discussed in the following section, $\Omega_{\chi_i}^2({\bf p};T)$ can
be well approximated as
\begin{eqnarray}
  \Omega_{\chi_i}^2({\bf p};T) = {\bf p}^2 + \tilde{m}_{\chi_i}^2 (T),
\end{eqnarray}
where $\tilde{m}_{\chi_i}^2 (T)$ is given by the sum of bare and
thermal masses.  

The effect of non-vanishing $\Gamma_{\chi_i}$ will be numerically
studied in the next section.  Here, we comment that, if
$\Gamma_{\chi_i}$ is small enough, the evolution of the total number
density of $\phi$ is governed by a simple differential equation.  If
the interaction is perturbative, it is usually the case that
$\Gamma_{\chi_i}({\bf p};T)\ll\tilde{m}_{\chi_i} (T)$.  In such a
case, the (Breit-Wigner) spectral density is well approximated as
$\rho_{\chi_i} (p^0,{\bf p}; T) \simeq 2\pi {\rm sign}(p^0)
\delta({p^0}^2-{\bf p}^2-\tilde{m}_{\chi_i}^2 (T))$, and hence
$G^{\chi_i}_>(\omega,{\bf k})$ and $G^{\chi_i}_<(\omega,{\bf k})$ are
also (approximately) proportional to the $\delta$-function.  We call
this limit as zero-width limit.  
Then, the collision term in Eq.\ \eqref{eq:Ncoll} becomes
\begin{eqnarray}
  \left[\dot{n}_\phi^{\rm (Coll)}\right]_{\Gamma_{\chi_i}\to 0} &=& 
  g^2 \int d\Pi_\phi^{(k^0>0)} (k) 
  \left[ \prod d\Pi_{\chi_i} (p_i) \right] 
  (2\pi)^4 
  \delta (k^0 - \mbox{$\sum_i$} p_i^0) 
  \delta^{(3)} ({\bf k}- \mbox{$\sum_i$} {\bf p}_i)
  \nonumber \\ &&
  \left[ \prod_i (1 + f_{\rm B} (p_i^0)) {\rm sign}(p_i^0)
    - \prod_i f_{\rm B} (p_i^0) {\rm sign}(p_i^0)
  \right]
  \left[ f_{\rm B} (k^0) - N_{\bf k} \right],
  \label{ndot_coll}
\end{eqnarray}
where
\begin{eqnarray}
  d\Pi_{\chi_i} (p_i) = 
  \frac{d^4 p_i}{(2\pi)^3} \delta (p_i^2 - \tilde{m}_{\chi_i}^2(T)),
\end{eqnarray}
and $d\Pi_\phi (k)$ is defined in the same way.  (Here, we have
introduced the four-component vector as $p_i=(p_i^0,{\bf p}_i)$.)
Notice that, in Eq.\ \eqref{ndot_coll}, the $p_i^0$ integration is
performed in the region $-\infty<p_i^0<\infty$, while $k^0$
integration is for $k^0>0$.

It is notable that the collision term given in Eq.\ \eqref{ndot_coll}
includes all the relevant scattering and decay processes (and their
inverse processes).  Regarding $\chi_i$ as scalar particles with masses
$\tilde{m}_{\chi_i}(T)$, the integrand of the collision term becomes
non-vanishing if (and only if) the momentum configurations are
kinematically allowed.  In addition, Eq.\ \eqref{ndot_coll} contains
the effect of induced emission.  For example, if the scattering process
$\chi_1(p_1)\cdots\chi_I(p_I)\leftrightarrow
\phi(k)\chi'_1(q_1)\cdots\chi'_F(q_F)$ is kinematically allowed, the
collision term contains
\begin{eqnarray}
  \left[\dot{n}_\phi^{\rm (Coll)}\right]_{\Gamma_{\chi_i}\to 0} 
  &\supset&
  g^2 \int
  \left[ \prod d\Pi_{\chi_i}^{(p^0>0)} (p_i) \right] 
  \left[ \prod d\Pi_{\chi'_f}^{(q^0>0)} (q_f) \right] 
  d\Pi_\phi^{(k^0>0)} (k) 
  \nonumber \\ &&
  (2\pi)^4 
  \delta (k^0 + \mbox{$\sum_f$} q_f^0 - \mbox{$\sum_i$} p_i^0) 
  \delta^{(3)} 
  ({\bf k}+\mbox{$\sum_f$} {\bf q}_f-\mbox{$\sum_i$} {\bf p}_i)
  \nonumber \\ &&
  \big\{ 
  [ \mbox{$\prod_i$} f_{\rm B} (p_i^0) ]
  [ \mbox{$\prod_f$} (1 + f_{\rm B} (q_f^0)) ]
  (1 + N_{\bf k})
  -  [ \mbox{$\prod_i$} (1+f_{\rm B} (p_i^0)) ]
  [ \mbox{$\prod_f$} f_{\rm B} (q_f^0) ]
  N_{\bf k} \big\},
  \nonumber \\
\end{eqnarray}
where we have used the relation $f_{\rm B}(-\omega)=-(1+f_{\rm
  B}(\omega))$.  The right-hand side of the above equation has the
same structure as the collision term in conventional Boltzmann
equation; however, notice that the thermally corrected dispersion
relations should be used in evaluating both the phase-spaces and
distribution functions.

\section{Application to non-thermal dark matter production}
\label{sec:dm}
\setcounter{equation}{0}

Now, let us apply the formalism to the non-thermal dark matter
production scenario, which is recently called freeze-in scenario,
regarding $\phi$ as dark matter.  In such a scenario, the dark matter
particle is always out of thermal equilibrium because of the weakness
of its interaction, and is produced by the decay of particles in
thermal bath, $\chi_i$.  Thus, this is the situation where we can
safely use the formalism discussed in the previous section.  We pay
particular attention to the question how large the thermal effect can
be.  For this purpose, we numerically calculate the production rate
and the relic abundance of $\phi$ using the formalism presented in the
previous section.

Here, we consider the simplest form of the interaction term, which is
\begin{eqnarray}
  {\cal L}_{\rm int}
  = g \phi {\cal O}[\chi_0,\chi_1]
  = g \phi\chi_0\chi_1.
  \label{L_int2}
\end{eqnarray}
For simplicity, we concentrate on the case that $\phi$, $\chi_0$, and
$\chi_1$ are all real scalars, with masses satisfying $m_{\chi_0} >
m_\phi + m_{\chi_1}$.  In our numerical calculations, we take
$m_{\chi_0}= 100\ {\rm GeV}$ and $m_{\chi_1}=0$.  Then, in the absence
of thermal effects, $\phi$ can be produced by the two-body decay
process $ \chi_0\rightarrow\phi\chi_1$.  As we will see in the
following, such a decay process may be blocked, or even the ``decay''
process $\chi_1\rightarrow\phi\chi_0$ may occur, once thermal effects
are taken into account.

At the leading-order in $g$, $\phi$ is produced only by
$1\leftrightarrow 2$ ``decay" processes.  Using the formulae given in
the previous section, we obtain
\begin{eqnarray}
  \Gamma_\phi ({\bf k}; T) =
  \frac{g^2}{2\omega_{\bf k}}
  \int \frac{d^4 q}{(2\pi)^4}
  \left[ 1 + f_{\rm B}(q_0) + f_{\rm B}(\omega_{\bf k}-q_0) \right]
  \rho_{\chi_0}
  (q_0, {\bf q})\rho_{\chi_1}(\omega_{\bf k}-q_0,{\bf k}-{\bf q}).
\end{eqnarray}
As we have discussed in the previous section, when the quasi-particle
picture is applicable, the thermal effects on $\chi_i$ are imprinted
in $\Omega_{\chi_i}$ and $\Gamma_{\chi_i}$.  Importantly, these
quantities depend on how $\chi_i$ interacts in thermal bath.  To make
our discussion concrete, we adopt the following form of the
interaction:\footnote{Although $\chi_i$ couples to other
  particles in the thermal bath, here we assume that the dominant effect
  on the dispersion relation of $\chi_i$ is from its self
  interaction.}
\begin{eqnarray}
  {\cal L}_{\rm int} = - \frac{1}{4!} g_{\chi_i}^2 \chi_i^4.
\end{eqnarray}
Then, at the leading order in $g_{\chi_i}$, the thermal mass of
$\chi_i$ is obtained as \cite{Dolan:1973qd}
\begin{eqnarray}
  \tilde{m}_{\chi_i}^2 = m_{\chi_i}^2 +
  \frac{g_{\chi_i}^2}{2}
  \int \fr{d^3 p}{(2\pi)^3} 
  \frac{f_{\rm B}(\omega_{\bf p}^{\chi_i}) }{\omega_{\bf p}^{\chi_i}},
  \label{thermalmass}
\end{eqnarray}
where $\omega_{\bf p}^{\chi_i}=\sqrt{m_{\chi_i}^2 + {\bf
    p}^2}$.  In the following analysis, we consider relatively large
value of $g_{\chi_1}$.  Thus, we include the NLO contribution to the
thermal mass of $\chi_1$.  In the present analysis, as we mentioned,
we take $m_{\chi_1}=0$. Then, the thermal mass of $\chi_1$ is given 
by, at NLO in $g_{\chi_1}$ \cite{Dolan:1973qd}
\begin{eqnarray}
  \tilde{m}_{\chi_1}^2 (T) = \fr{1}{24} g_{\chi_1}^2 T^2 
    \( 1 - \fr{3}{\pi } \fr{g_{\chi_1}}{\sqrt{24}} \).
    \label{eq:thermalmass_NLO}
\end{eqnarray}
Note that we drop the momentum dependence, since, at this order, the
only contribution is a tad pole diagram.  In addition,
$\Gamma_{\chi_i}$ is given in the integral form as~\cite{Hosoya:1983ke+x}
\begin{eqnarray}
  \Gamma_{\chi_i}({\bf p};T) &=& 
  \frac{\pi}{2\Omega_{\chi_i} ({\bf p})} g_{\chi_i}^4
  \int \frac{d^3 q_1}{(2\pi)^32\Omega_{\chi_i,1}}
  \int \frac{d^3  q_2}{(2\pi)^32\Omega_{\chi_i,2}}
  \int \frac{d^3  q_3}{(2\pi)^32\Omega_{\chi_i,3}}
  \nonumber \\ &&
  (2\pi)^3 \delta({\bf p}+{\bf q}_1-{\bf q}_2-{\bf q}_3)
  \delta (\Omega_{\chi_i} ({\bf p}) + \Omega_{\chi_i,1}
  - \Omega_{\chi_i,2} - \Omega_{\chi_i,3})
  \nonumber \\ &&
   \left[ 
     f_{{\rm B},1}(1 + f_{{\rm B},2}) (1 + f_{{\rm B},3})
     - (1 + f_{{\rm B},1}) f_{{\rm B},2} f_{{\rm B},3}
   \right],
\end{eqnarray}
where, for the simplicity of the equation, we defined $\Omega_{\chi_i,
  J}\equiv\Omega_{\chi_i} ({\bf q}_J)$ 
and $f_{{\rm B},J}\equiv f_{\rm B} (\Omega_{\chi_i,J})$
(with $J=1-3$).

With the above formulae, we can follow the evolution of the number
density of $\phi$ in the early universe and calculate the relic
abundance.  For this purpose, it is convenient to define the ``yield
variable'' as
\begin{eqnarray}
  Y_\phi \equiv \frac{n_\phi}{s},
\end{eqnarray}
where $s=\frac{2\pi^2}{45}g_*T^3$ is the entropy density.  (Here,
$g_*$ is the effective number of relativistic degrees of freedom; in
our numerical calculation, we use $g_*=100$.)   We are interested in
the case that $N_{\bf k}\ll f_{\rm B} (\omega_{\bf k})$ and, in such a
case, we can neglect the term proportional to $N_{\bf k}$ in the
right-hand side of Eq.\ \eqref{eq:Ncoll}.  Then, the present value
of $Y_\phi$ is given by
\begin{eqnarray}
  Y_\phi (T) \equiv 
  \int_{-\infty}^{\log(m_{\chi_0}/T)}
  d\log z \frac{d Y_\phi}{d \log z},
\end{eqnarray}
where
\begin{eqnarray}
  \frac{d Y_\phi}{d \log z} = 
  \left. \frac{\dot{n}_\phi^{\rm (Coll)}}{sH} 
  \right|_{N_{\bf k}\rightarrow 0}
  =
  \int \frac{d^3 k}{(2\pi)^3} 
  \frac{\Gamma_\phi ({\bf k};T) f_{\rm B} (\omega_{\bf k})}{sH},
  \label{dY/dlnT}
\end{eqnarray}
with
\begin{eqnarray}
  z \equiv \frac{m_{\chi_0}}{T}.
\end{eqnarray}
Notice that the variable $z$ increases as the universe expands.  We
will show how 
$dY_\phi/d\log z$ and $Y_\phi$ behave as a function of the cosmic
temperature $T$ in the following.

Before showing the numerical results, it is instructive to consider
the zero-width limit; in the present case, Eq.\ \eqref{ndot_coll}
becomes
\begin{eqnarray}
  \left[ \dot{n}_\phi^{\rm (Coll)} \right]_{\Gamma_{\chi_i}\rightarrow 0}
  &=& 
  \theta (\tilde{m}_{\chi_0} - m_\phi - \tilde{m}_{\chi_1})
  \int \frac{d^3 p}{(2\pi)^3} 
  \frac{\tilde{m}_{\chi_0}}{\Omega_{\chi_0}({\bf p})}
  \tilde{\Gamma}_{\chi_0\rightarrow\phi\chi_1} 
  f_{\rm B} (\Omega_{\chi_0}({\bf p}))
  [ 1 + \overline{f}_{\chi_1} ({\bf p}) ]
  \nonumber \\ && + 
  \theta (\tilde{m}_{\chi_1} - m_\phi - \tilde{m}_{\chi_0})
  \int \frac{d^3 p}{(2\pi)^3} 
  \frac{\tilde{m}_{\chi_1}}{\Omega_{\chi_1}({\bf p})}
  \tilde{\Gamma}_{\chi_1\rightarrow\phi\chi_0} 
  f_{\rm B} (\Omega_{\chi_1}({\bf p}))
  [ 1 + \overline{f}_{\chi_0} ({\bf p}) ],
  \nonumber \\
  \label{ndotc_zerowidth_twobody}
\end{eqnarray}
where we have used the relation $N_{\bf k}\ll f_{\rm B} (\omega_{\bf
  k})$.  In the above expression,
$\tilde{\Gamma}_{\chi_0\rightarrow\phi\chi_1}$ and
$\tilde{\Gamma}_{\chi_1\rightarrow\phi\chi_0}$ are decay rates of
$\chi_0$ and $\chi_1$ calculated with the conventional Feynman rules
but with thermally corrected dispersion relation; for example,
$\tilde{\Gamma}_{\chi_0\rightarrow\phi\chi_1}$ is given by
\begin{eqnarray}
  \tilde{\Gamma}_{\chi_0\rightarrow\phi\chi_1} = 
  \frac{g^2}{16\pi\tilde{m}_{\chi_0}}
  \sqrt{ 1 
    - \frac{2(m_\phi^2 + \tilde{m}_{\chi_1}^2)}{\tilde{m}_{\chi_0}^2}
    + \frac{(m_\phi^2 - \tilde{m}_{\chi_1}^2)^2}{\tilde{m}_{\chi_0}^4} }.
\end{eqnarray}
In addition, $\overline{f}_{\chi_0}$ and $\overline{f}_{\chi_1}$ are
``averaged'' distribution functions of $\chi_0$ and $\chi_1$ in thermal
bath, respectively.  For example, $\overline{f}_{\chi_0}$ is given by
\begin{eqnarray}
  \overline{f}_{\chi_0} ({\bf p}) = \frac{1}{2} \int d\cos\theta_{\rm CM}
  f_{\rm B} (\tilde{E}_{\chi_1}(|{\bf p}|,\theta_{\rm CM})),
\end{eqnarray}
where $\tilde{E}_{\chi_1}(|{\bf p}|,\theta_{\rm CM})$ denotes the energy of
$\chi_1$ emitted from $\chi_0$ (carrying momentum $|{\bf p}|$) to the
direction $\theta_{\rm CM}$ relative to ${\bf p}$ (with $\theta_{\rm
  CM}$ being defined in the rest frame of $\chi_0$).  Notice that, in
the calculation of $\tilde{E}_{\chi_1}(|{\bf p}|,\theta_{\rm CM})$, dispersion
relations including the thermal masses should be used.  Eq.\
\eqref{ndotc_zerowidth_twobody} indicates that the production process
of $\phi$ is active only when $\tilde{m}_{\chi_0}
(T)>m_\phi+\tilde{m}_{\chi_1} (T)$ or $\tilde{m}_{\chi_1}
(T)>m_\phi+\tilde{m}_{\chi_0} (T)$;\footnote
{In the present case, $\phi$ should play the role of dark matter, and
  hence is stable.  Thus, we do not have to consider the case that
  $m_\phi>\tilde{m}_{\chi_0} (T)+\tilde{m}_{\chi_0} (T)\geq
  m_{\chi_0}+m_{\chi_0}$.}
otherwise, the production of $\phi$ is kinematically suppressed due
to the thermal mass.  As we will see below, this is indeed the case.

\begin{figure}[t]
\begin{center}
\subfigure{
\includegraphics[width=0.48\columnwidth]{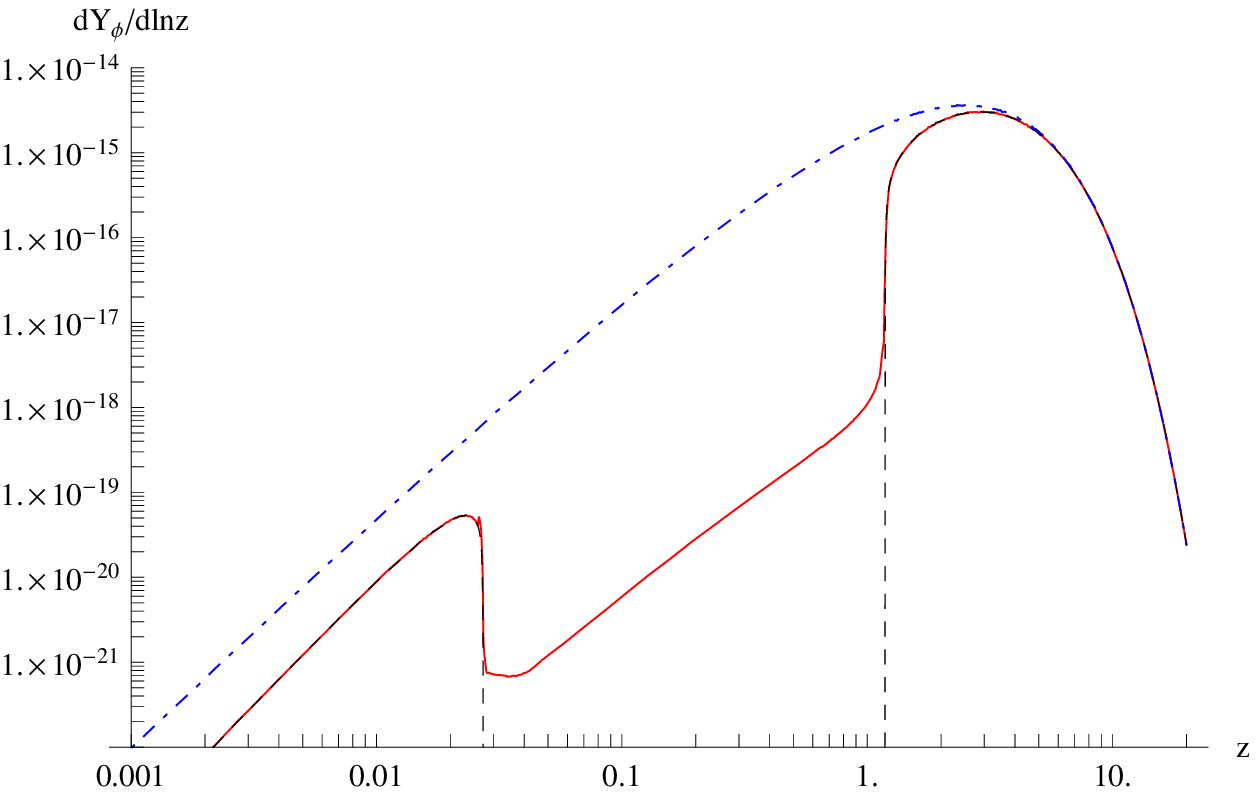}}
\subfigure{
\includegraphics[width=0.48\columnwidth]{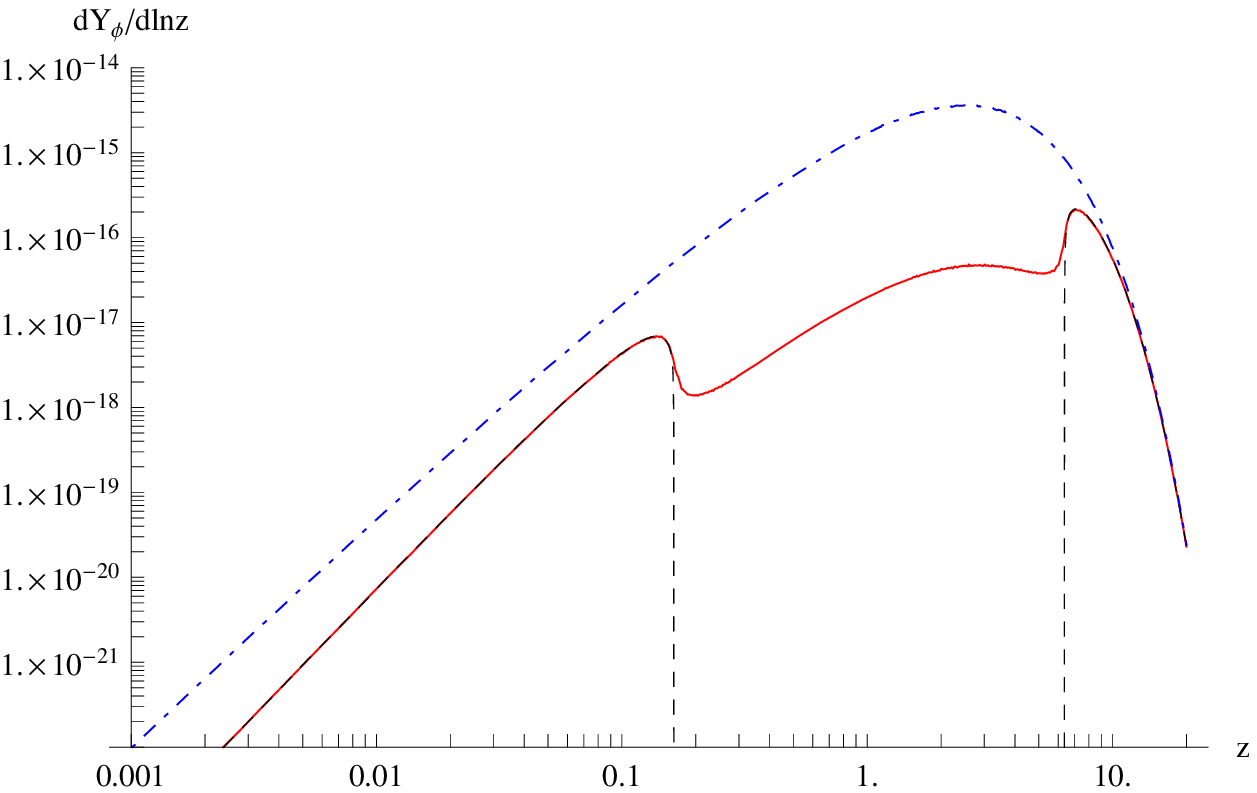}}
\caption{\small $dY_\phi/d\log z$ as a function of $z=m_{\chi_0}/T$
  for $m_\phi/m_{\chi_0}=0.95$, $g_{\chi_0}=0.1$,
  $g/m_{\chi_0}=10^{-13}$, and $g_*=100$.  The left figure is computed
  with $g_{\chi_1} = 0.3$ and the right one is with $g_{\chi_1} = 2$.
  The red solid lines are the results of full calculation while the
  black dashed ones are zero-width results.  Results of conventional
  Boltzmann equation are also shown in blue dot-dashed lines.}
\label{fig:dY/dlnz}
\end{center}
\end{figure}

Now we are at the position to numerically calculate the production
rate of $\phi$ in thermal bath.  First, let us discuss how
$dY_\phi/d\log z$ given in Eq.\ \eqref{dY/dlnT} behaves.  In Fig.\
\ref{fig:dY/dlnz}, we plot $dY_\phi/d\log z$ as a function of
$z=m_{\chi_0}/T$.  Here we take $m_\phi/m_{\chi_0}=0.95$, $g_{\chi_0}
= 0.1$, and $g_{\chi_1} = 0.3$ (left) and $g_{\chi_1}=2$ (right).  In
each figure, the result of the full calculation, which is calculated
by using Eq.\ \eqref{dY/dlnT} including the effects of the widths of
$\chi_i$, $\Gamma_{\chi_i}$, is shown in the red solid line.  In the
same figure, in the black dashed line, we show the result taking the
zero-width approximation (see Eq.\ \eqref{ndotc_zerowidth_twobody}).
The result of the calculation with conventional Boltzmann equation
(which corresponds to the zero-width approximation with
$\tilde{m}_{\chi_i}\rightarrow m_{\chi_i}$) is also shown in the blue
dot-dashed line.

As one can see, $dY_\phi/d\log z$ is significantly suppressed when
$0.027 \lesssim m_{\chi_0}/T \lesssim 1.2$ (for $g_{\chi_1} = 0.3$) or
$0.16 \lesssim m_{\chi_0}/T \lesssim 6.3$ (for $g_{\chi_1} = 2$). This
is due to the fact that, at leading order in
$\Gamma_{\chi_i}/\tilde{m}_{\chi_i}$, the production process of $\phi$
is kinematically blocked because of the thermal mass; the upper and
lower edges correspond to the temperature where
$\tilde{m}_{\chi_1}\simeq m_\phi + \tilde{m}_{\chi_0}$ and
$\tilde{m}_{\chi_0}\simeq m_\phi+\tilde{m}_{\chi_1}$ are realized,
respectively.  In the suppressed region between the edges, the $\phi$
production is from the off-shell effects of $\chi_i$ (especially
$\chi_1$), namely due to non-vanishing $\Gamma_{\chi_i}$
\cite{Yokoyama:2005dv,Drewes:2010pf}.

It is notable that, in the region where there is no kinematical
suppression, the result of zero-width approximation agrees remarkably
well with that of full calculation.  
Thus, the effect of $\Gamma_{\chi_i}$ can be safely neglected in the
calculation of the relic abundance of $\phi$, unless the production
channel is almost kinematically blocked at the epoch when the
production of $\phi$ is most active. If the zero-width approximation
can be adopted, the calculation of the production rate is extremely
simplified.  In the opposite case where $\phi$ is mostly produced at
the epoch of kinematical suppression, production processes due to
$\Gamma_{\chi_i} \neq 0$ may not be negligible in the calculation of
the relic abundance of $\phi$.  A typical example is given in the
right hand side of Fig.\ \ref{fig:dY/dlnz}.

From Fig.\ \ref{fig:dY/dlnz}, we can also see that, if we perform a
naive calculation without properly taking account of the thermal
effect, the production rate may significantly deviate from the correct
value.  In particular, even at $\tilde{m}_{\chi_1} \gtrsim m_\phi +
\tilde{m}_{\chi_0}$ or $\tilde{m}_{\chi_0} \gtrsim m_\phi +
\tilde{m}_{\chi_1}$, the discrepancy between the results of full and
conventional calculations is sizable in some parameter region.  This
is because, at $\tilde{m}_{\chi_1} \gtrsim m_\phi +
\tilde{m}_{\chi_0}$, the dominant production process of $\phi$ is
$\chi_1\rightarrow\phi\chi_0$ contrary to the case of low enough
temperature, and at $\tilde{m}_{\chi_0} \gtrsim m_\phi +
\tilde{m}_{\chi_1}$, the phase space becomes smaller than that in the
case of conventional calculation due to the thermal mass of $\chi_1$.
\begin{figure}[t]
  \begin{center}
    \subfigure{
      \includegraphics[clip,width=0.48\columnwidth]{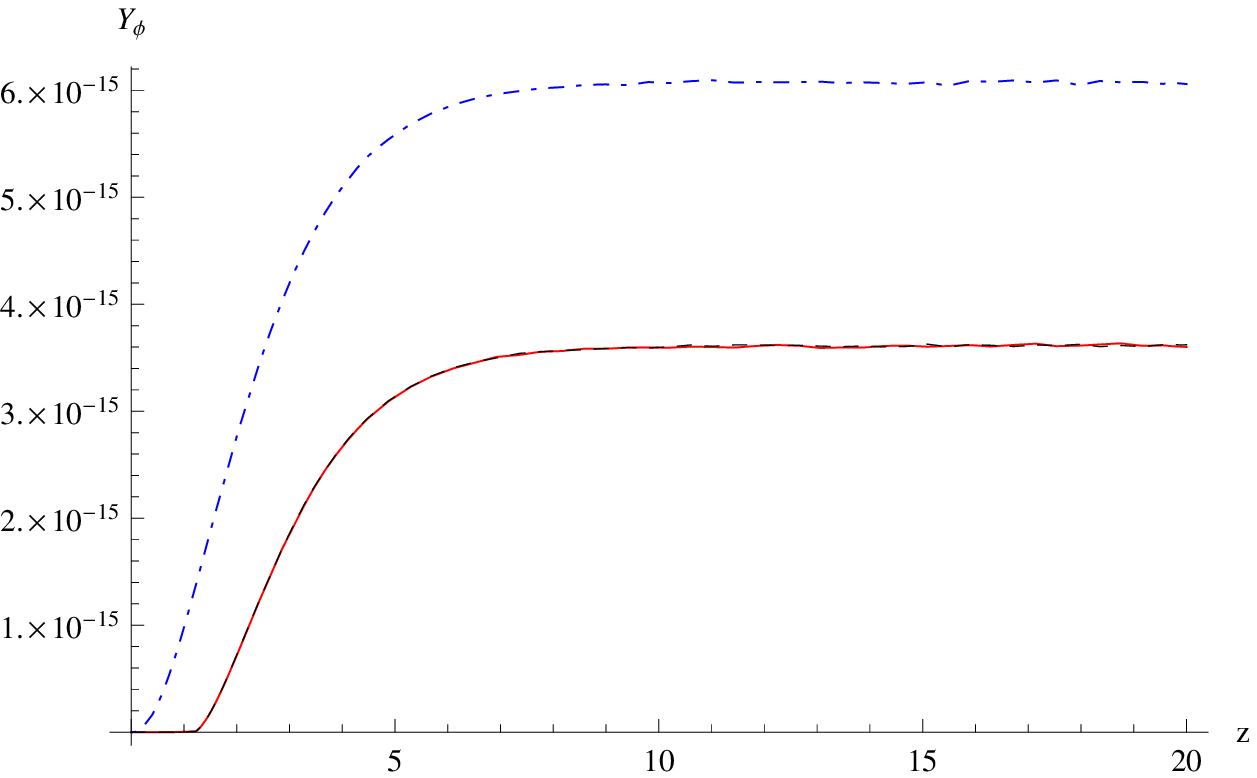}}
    \subfigure{
      \includegraphics[clip,width=0.48\columnwidth]{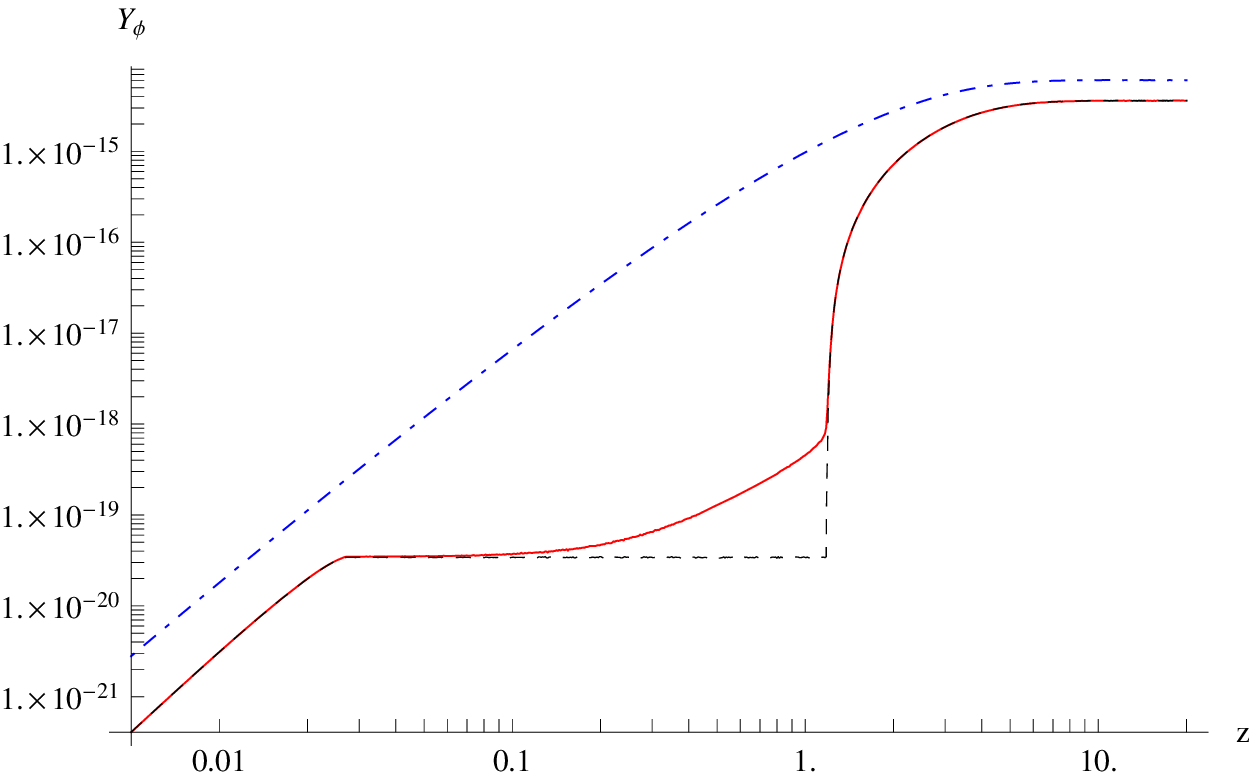}}
    \caption{\small The evolution of the yield variable $Y_\phi$ for
      $m_\phi/m_{\chi_0}=0.95$, $g_{\chi_0}=0.1$, $g_{\chi_1}=0.3$,
      $g/m_{\chi_0}=10^{-13}$, and $g_*=100$. The red solid  line is
      computed with $\Gamma_{\chi_i} \ne 0$ and the black dashed one is done
      with $\Gamma_{\chi_i} = 0$.  The blue dot-dashed one represents the
      conventional result (without thermal effects).}
    \label{fig:abundance03}
  \end{center}
  \begin{center}
    \subfigure{
      \includegraphics[clip,width=0.48\columnwidth]{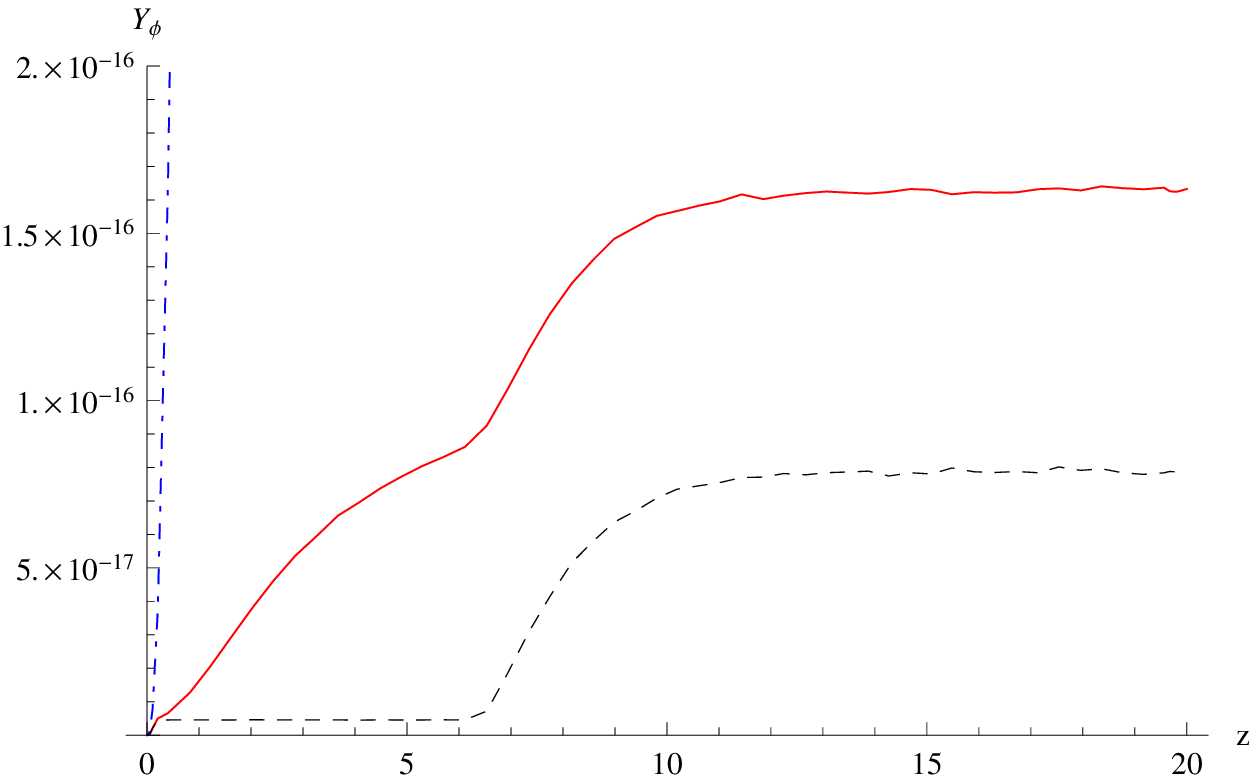}}
    \subfigure{
      \includegraphics[clip,width=0.48\columnwidth]{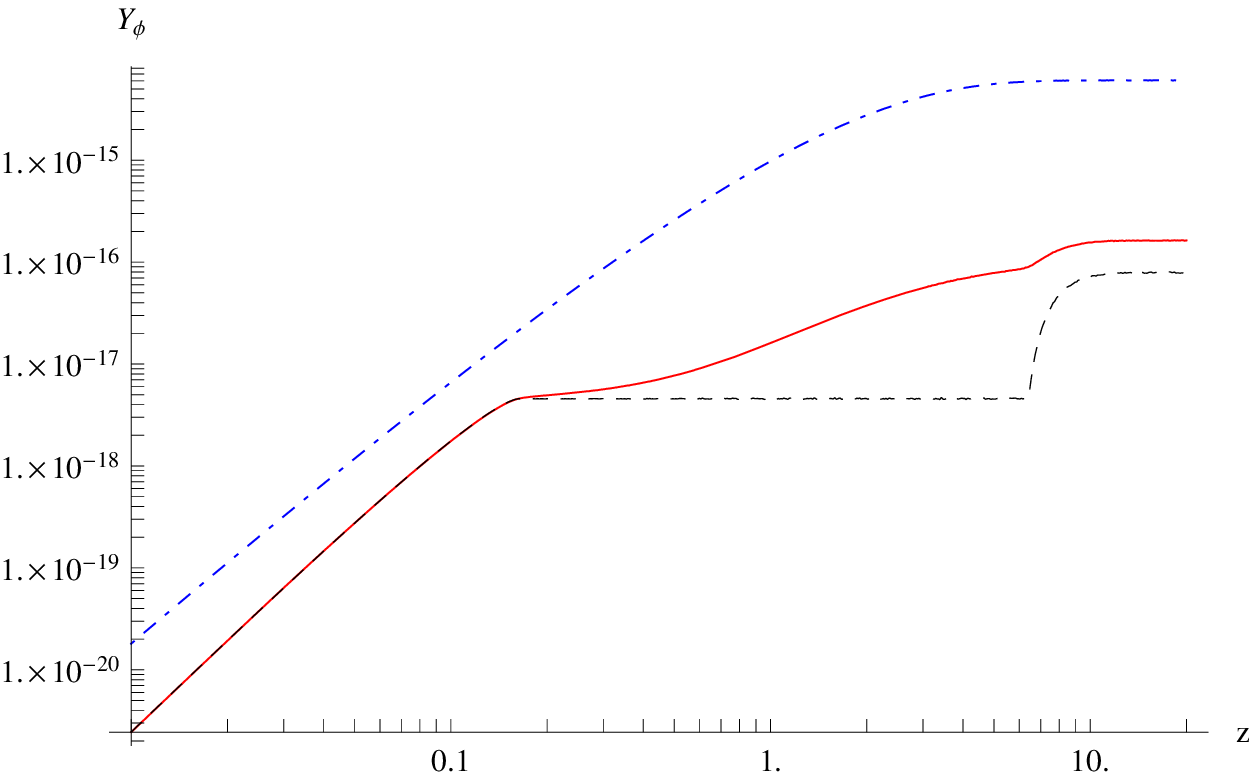}}
    \caption{\small Same as Fig.\ \ref{fig:abundance03} except for
      $g_{\chi_1}=2$. Notice that, in the left figure, the line for
      the conventional result is almost parallel to the vertical
      axis.}
    \label{fig:abundance20}
  \end{center}
\end{figure}

In the non-thermal dark matter production due to the decay, the dominant
production occurs when $m_{\chi_0}/T \simeq 1-5$; for such a
temperature, the actual value of $dY_\phi/d\log z$ may receive sizable
thermal effects.  Thus, for the accurate calculation of the relic
density of non-thermally produced dark matter, it is dangerous to
neglect the thermal effects.
In Figs.\ \ref{fig:abundance03} and \ref{fig:abundance20}, we show the
evolution of the yield variable $Y_\phi$.  In the same figure, we also
show the result of the naive calculation neglecting the thermal
effects.  As one can see, the evolution of $Y_\phi$ changes once the
thermal effects are taken into account, and the resultant value of
$Y_\phi$ is suppressed compared to the result obtained by neglecting
the thermal effects.  In particular, for the case of $g_{\chi_1} = 2$,
which gives relatively large value of $\Gamma_{\chi_1}$, significant
amount of $\phi$ is produced even during the period of kinematical
suppression and the resultant $Y_\phi$ differs from the zero-width
result.  This is because the production channel is suppressed during
the time $m_{\chi_0}/T\simeq 1-5$, where the most effective production
could occur if there were no thermal effects.  Therefore, the
production due to the off-shell effects with $\Gamma_{\chi_i}\neq 0$
becomes important.  In such a case, the resultant value of $Y_\phi$
does not agree with the result of zero-width approximation nor that of
the conventional Boltzmann equation.

\begin{figure}[t]
\begin{center}
\includegraphics[width=10cm]{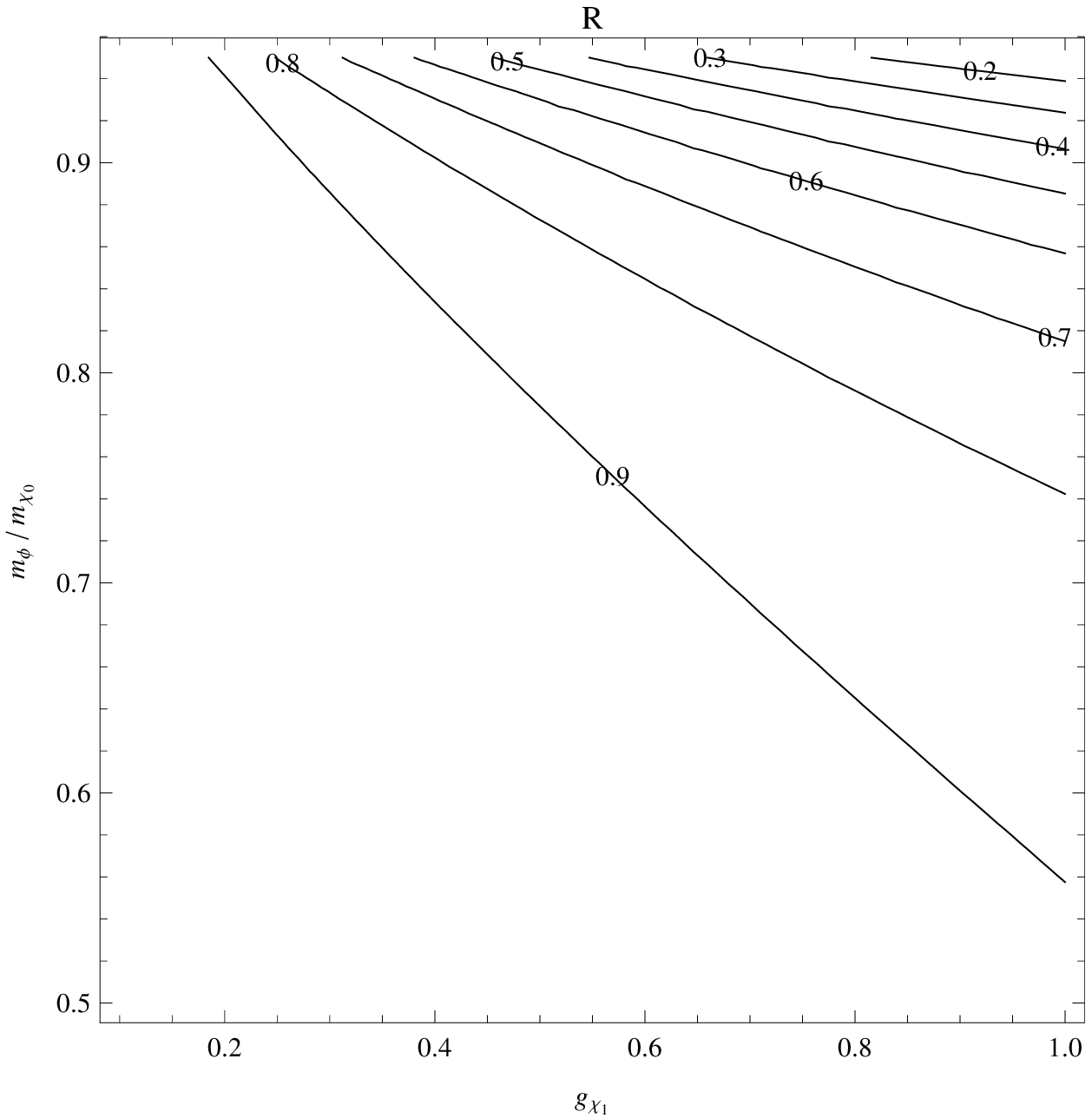}
\caption{\small The contour plot of $R$ as a function of $g_{\chi_1}$
  and $m_\phi/m_{\chi_0}$, taking $g_{\chi_0}=0.1$. 
 }
\label{fig:ratio}
\end{center}
\end{figure}

In Fig.\ \ref{fig:ratio}, we plot the ratio of the yield variable
calculated with and without thermal effects,
\begin{eqnarray}
  R \equiv 
  \left. 
    \frac{Y_\phi}{Y_\phi (\mbox{conventional Boltzmann equation})} 
  \right|_{\rm now},
\end{eqnarray}
as a function of $g_{\chi_1}$ and $m_\phi/m_{\chi_0}$, 
using the zero-width approximation.
In the present
set up, the present number density of $\phi$ decreases by taking
account of the thermal effects.  As one can see, $Y_\phi$ is more
suppressed when $g_{\chi_1}$ becomes larger or when the mass of $\phi$
becomes closer to that of $\chi_0$.

Before closing this section, we note here that there is another
possible mechanism of enhancing the production rate.  With the
emission of a massless particle in the thermal bath from the initial
or final state particles (corresponding to $\chi_0$ or $\chi_1$ in the
present setup), the production rate of the non-thermal dark matter at
$z\lesssim 1$ may be affected, and may become larger by the factor of
a few \cite{Anisimov:2010gy}.  In realistic models of non-thermally
produced dark matter, the particles corresponding to $\chi_0$ and
$\chi_1$ may couple to massless gauge bosons, which may enhance the
production rate at $z\lesssim 1$.  In the present model, however,
there is no massless particle responsible for such an enhancement.  In
addition, for the calculation of the relic density of the
non-thermally produced dark matter, such an effect may not be
important because, as we have shown, the relic abundance of the
non-thermally produced dark matter is determined at $z\sim O(1)$.

\section{Conclusions and Discussion}
\label{sec:conclusions}
\setcounter{equation}{0}

In this paper, we have discussed the evolution of the number density
of a particle $\phi$ which is coupled to thermal bath very weakly and
hence is in non-equilibrium.  We first solved the
Kadanoff-Baym equations for the case that (i) the effects of $\phi$ on
the thermal bath is negligible and (ii) the self interaction of $\phi$
is sufficiently weak.  Then, we derived Boltzmann equation describing
the evolution of the $\phi$'s number density for the case that (iii)
the real part of the $\phi$'s self energy is (almost) unchanged by
thermal effects.  We then studied the properties of Boltzmann
equation, assuming that the degrees of freedom in the thermal bath are
well described as the quasi-particles.  In particular, in the
situation that the widths of quasi-particles are negligible, the
evolution of the number density is well described by the Boltzmann
equation in the familiar form, which contains the matrix elements,
phase-space integrals, and distribution functions evaluated with the
``on-shell'' condition modified by the thermal effects.

Then, we have applied the formalism to the scenario in which dark
matter is non-thermally produced from the decay of ``particles'' in
thermal bath, regarding $\phi$ as the dark matter.  In such a case,
the above conditions (i), (ii), and (iii) are satisfied, and the
formalism we have studied can be safely applied.  We calculated the
number density of the non-thermally produced dark matter, taking
account of the effects of thermal bath.  Because of the change of the
dispersion relations of ``particles'' in thermal bath, the production
rate of the dark matter may significantly change.  In particular, in
some cases, the decay process to produce $\phi$, which is
kinematically allowed in the vacuum, may be blocked because the
particles in thermal bath acquire thermal masses, which changes the
mass relation among parent and daughter quasi-particles.  Numerically,
we found that, if the proper Boltzmann equation with the thermal
effects is used, the dark matter density may change by $O(10-100\ \%)$
compared to the results of calculations neglecting the change of the
dispersion relation of the ``particles'' in thermal bath.  We have
also studied the effect of the width in the spectral density of the
quasi-particles in thermal bath, and found that the zero-width limit
can be safely taken if the thermal blocking of the production of
$\phi$ does not occur at the time when the production of $\phi$ is
most effective.

\section*{Acknowledgements}
We would like to thank M.~Endo for collaboration at an early stage of this work.
This work is supported by Grant-in-Aid for Scientific research from
the Ministry of Education, Science, Sports, and Culture (MEXT), Japan,
No.\ 22244021 (K.H. and T.M.), No.\ 21740164 (K.H.), and No.\
22540263 (T.M.), and by World Premier International Research Center
Initiative (WPI Initiative), MEXT, Japan.

\end{document}